\documentclass[twocolumn,floatfix,showpacs,aps,nofootinbib]{revtex4}
\usepackage{color}
\usepackage{amsmath}

\newcommand {\bea}{\begin{eqnarray}}
\newcommand {\eea}{\end{eqnarray}}
\newcommand {\be}{\begin{equation}}
\newcommand {\ee}{\end{equation}}

\begin{document}

\def\Journal#1#2#3#4{{#1} {\bf #2}, #3 (#4)}
\def\PRC{{Phys. Rev. C}}
\def\PRD{{Phys. Rev. D}}
\def\FP{{Foundations of Physics}}
\def\ZPA{{Z. Phys. A}}
\def\NPA{{Nucl. Phys. A}}
\def\JPG{{J. Phys. G Nucl. Part}}
\def\PRL{{Phys. Rev. Lett}}
\def\PRpt{{Phys. Rep.}}
\def\PLB{{Phys. Lett. B}}
\def\AP{{Ann. Phys (N.Y.)}}
\def\EPJA{{Eur. Phys. J. A}}
\def\NP{{Nucl. Phys}}  
\def\ZP{{Z. Phys}}
\def\RMP{{Rev. Mod. Phys}}
\def\IJMPE{{Int. J. Mod. Phys. E}}

\title{From self-consistent covariant effective field theories to their Galilean-invariant counterparts}

\author{A. Sulaksono$^{1,5}$, P. -G. Reinhard$^{2}$, T. J. B{\"u}rvenich$^{3}$, P. O. Hess$^{3,4}$ and J. A. Maruhn$^{5}$}

\affiliation{$^1$Departemen Fisika, FMIPA, Universitas Indonesia,
Depok, 16424, Indonesia\\
$^2$Institut
f\"ur Theoretische Physik II, Universit\"at Erlangen-N\"urnberg, D-91058 Erlangen, Germany\\
$^3$Frankfurt Institute for Advanced Studies, Universit\"at Frankfurt, 60438 Frankfurt am Main, Germany\\
$^4$Instituto de Ciencias Nucleares, Universidad Nacional Aut\'onoma de M\'exico, Apdo. Postal 70-543, M\'exico 04510 D.F\\
$^5$Institut
f\"ur Theoretische Physik, Universit\"at Frankfurt, 60438 Frankfurt am Main, Germany}

\begin{abstract}
We discuss how to obtain the nonrelativistic limit of a
self-consistent relativistic effective field theory for dynamic
problems.  It is shown that the standard $v/c$ expansions yields
Galilean invariance only to first order in $v/c$, whereas second order
is required to obtain important contributions such as the spin-orbit
force.  We propose a modified procedure which is a mapping rather than
a strict $v/c$ expansion.
\end{abstract}
\maketitle

\section{Introduction}
\label{sec_intro}
Since the development of the special theory of relativity classical
physics is viewed as its {\it nonrelativistic limit\/} for $c$, the
velocity of light, going to infinity (or correspondingly, considering
all relevant velocities to be much smaller than $c$). In Quantum
Mechanics, Dirac's theory of the electron carrying spin 1/2 has as its
nonrelativistic limit the Pauli equation.

Since then, the problem in obtaining a nonrelativistic limit of a
theory has enjoyed permanent interest. Many physical systems are on
the borderline between relativistic and classical physics, determined
by the velocity and/or the size of the system.  Prominent examples are
heavy atoms and nuclei, where the spin-orbit force is a relativistic
effect and shows that, although a nonrelativistic description is in
general easier to handle, relativistic effects cannot always be
neglected. Understanding the transition from relativistic to
nonrelativistic physics while maintaining the traces of relativistic
effects, is therefore of utmost interest.

One of the first, and still most prominent, attempts to handle this
transition in a systematic way was the Foldy-Wouthysen transformation
\cite{foldy} using a canonical transformation to obtain two equations
with two components, one of which becomes the Pauli equation in the
nonrelativistic limit.  An alternative method is related to group
contractions
\cite{gilmore},
where one discusses under which conditions a group can be {\it
contracted} to another one, involving non singular transformations.
In particular, the Lorentz group $SO(3,1)$ can be contracted to the
Galilei group, taking the limit $c\rightarrow\infty$ and neglecting
corrections of the order of $(\frac{v}{c})^2$.  On the level of group
generators this contraction procedure is direct, but on that of
representations it is not as trivial
\cite{gilmore}.
As we shall see, this is the case in
self-consistent effective field theories involving fields with a spin $\frac{1}{2}$
representation.
In recent years, several groups discussed similar problems
\cite{montigny,holland1,holland2} for the Maxwell and Dirac
equations. They showed that several nonrelativistic limits may exist
in these cases.

In the present paper, we focus on the nonrelativistic limit of
covariant nonlinear self-consistent effective field theories, i.e. those
based on density functionals.
These cases are to be distinguished from
other effective field theories which rely on a straighforward expansion (see e.g.
 \cite{weinberg1}) and for which the following  consideration do not apply.
Realizations of self-consistent field theories 
include e.g.
the model by Duerr \cite{duerr}, Heisenberg's nonlinear spinor theory
\cite{heisenberg}, the Nambu-Jona-Lasinio (NJL) model \cite{njl}, and
effective $\phi^4$ theories \cite{phi4}. In nuclear physics, the
relativistic mean-field (RMF) model \cite{reinhard,Ben03aR} and the
Skyrme-Hartree-Fock (SHF) approach \cite{Ben03aR} are typical
examples.

In this manuscript, we will discuss the link between such a
non-linear relativistic theory and its non-relativistic counterpart.
It is found that a straightforward nonrelativistic reduction of a
covariant ansatz up to $(v/c)^2$ yields a result which violates
Galileian invariance.  We will develop and justify a non-relativistic
mapping going up to $(v/c)^2$ that leads to the correct
Galilean-invariant counterpart.

\section{Nonrelativistic reduction}
\label{sec_nrreduc}
\subsection{The goal}
A relativistic effective field theory expresses the configuration in
terms of Dirac four-spinor wavefunctions $\psi_\alpha$ for each
state $\alpha$.  For the further developments, it is useful to
express it explicitly through upper and lower two-spinor
components as
\begin{equation}
  \psi_\alpha
  =
  \left(\begin{array}{c}
    \varphi^{\rm(u)}_\alpha \\
    \varphi^{\rm(d)}_\alpha
  \end{array}\right).
\label{eq:expa-wf}
\end{equation}
For means of simplicity and lucidity we choose a simple covariant self-consistent effective
field theory involving point couplings between its degrees of freedom, i.e., the
4-component spinors. It reads
\begin{eqnarray}\label{eq:prmf_den}
  {\cal L}_{{\rm c}} 
  &=&{\cal L}_{\rm free}+ 
  \frac{c_s}{2}\varrho_s^{2} +  
  \frac{c_s}{2}\varrho_\mu^{\mbox{}}\varrho^\mu_{\mbox{}},
\end{eqnarray}
with
\begin{subequations}
\begin{eqnarray}
  \varrho_s
  &=& 
  \sum_\alpha\bar\psi_\alpha\psi_\alpha 
  =
  \sum_\alpha \left[
   \varphi^{{\rm(u)}\dag}_\alpha\varphi^{\rm(u)}_\alpha
   -
   \varphi^{{\rm(d)}\dag}_\alpha\varphi^{\rm(d)}_\alpha
  \right]
  \;,
\\
  \varrho_0
  &=&
  \sum_\alpha\bar\psi_\alpha\gamma_0\psi_\alpha
  =
  \sum_\alpha \left[
   \varphi^{{\rm(u)}\dag}_\alpha\varphi^{\rm(u)}_\alpha
   +
   \varphi^{{\rm(d)}\dag}_\alpha\varphi^{\rm(d)}_\alpha
  \right]
  ,
\\
  \mbox{\boldmath$\varrho$}
  &=&
  \sum_\alpha\bar\psi_\alpha\mbox{\boldmath$\gamma$}\psi_\alpha
  =
  \sum_\alpha \left[
   \varphi^{{\rm(u)}\dag}_\alpha\mbox{\boldmath$\sigma$}\varphi^{\rm(d)}_\alpha
   +
   \varphi^{{\rm(d)}\dag}_\alpha\mbox{\boldmath$\sigma$}\varphi^{\rm(u)}_\alpha
  \right]\quad
\end{eqnarray}
\label{eq:relat-dens}
\end{subequations}
with $\gamma_\mu=(\gamma_0,\mbox{\boldmath$\gamma$})$ the four-vector of Dirac matrices
\cite{Zuber}.
The relativistic functional appears simple because the kinetic and spin-orbit terms
are implicit in the scalar and vector densities, as we shall see.  For simplicity of notation, we
will drop the index label $\alpha$ in the following, identifying, e.g.,  the scalar
density with $\varrho_s=\bar\psi\psi$ and similarly for all other densities and
currents.

A key point in such approaches is, of course, correct normalization of 
the densities and consequently the wavefunctions. This implies
\begin{equation}
  \int {\rm d}^3r \bar\psi\gamma_0\psi
  =
  1
\label{eq:normcond}
\end{equation}
to guarantee invariance under Lorentz transformations. This may be surprising, since it involves the
zeroth component of a four vector, but is inevitable to counter the relativistic contraction of the
purely spatial volume element ${\rm d}^3r$ \cite{Zuber}. We will see that this is the key problem with
straightforward expansions and at the same time the key to the solution.

The goal is now to obtain a nonrelativistic functional based on the
densities (\ref{eq:ske_den}) from the relativistic parent functional
(\ref{eq:prmf_den}) in a nonrelativistic limit.  Solutions in the
positive-energy branch (particle-like) are distinguished by a
dominance of $\varphi^{\rm(u)}$ over $\varphi^{\rm(d)}$. The strategy
is thus to eliminate $\varphi^{\rm(d)}$ and identify the upper
component with the classical two-spinor wavefunction,
$\varphi^{\rm(u)}\longleftrightarrow\varphi^{\rm(cl)}$, which should
obey the normalization condition
\begin{equation}
  \int {\rm d}^3r\bigl|\varphi^{\rm(cl)}\bigr|^2=1.
\label{eq:clnormcond}
\end{equation}
We can thus express $\psi$ in terms of $\varphi^{\rm(cl)}$.  Inserting that
into the coupling Lagrangian density (\ref{eq:prmf_den}) should produce the desired limit.

In the course of obtaining the nonrelativistic limit we expect to formulate the
nonrelativistic counterpart in terms of the following densities and currents:
\begin{subequations}
\label{eq:ske_den}
\begin{eqnarray}
 \rho 
 &=& 
 \sum_\alpha\left|\varphi^{\rm cl}_\alpha\right|^2 ,
 \quad
 \tau
 \;=\; 
 \sum_\alpha \left|\nabla\varphi^{\rm cl}_\alpha\right|^2 ,
\\ 
 {\bf J} 
 &=& 
 -\frac{i}{2} \sum_\alpha [\varphi^{{\rm cl}\dag}_\alpha 
  (\nabla\!\times\!\sigma)\varphi^{\rm cl} -
  {(\nabla\!\times\!\sigma\varphi^{\rm cl}_\alpha )}^{\dag}
 \varphi^{\rm cl}_\alpha] , 
\\ 
 {\bf j} 
 &=& 
 -\frac{i}{2}\sum_\alpha
 [\varphi^{{\rm cl} \dag}_\alpha {\nabla}\varphi^{\rm cl}_\alpha
  -({\nabla} \varphi^{\rm cl}_\alpha )^{\dagger} 
 \varphi^{\rm cl}_\alpha] , 
\\ 
 \mbox{\boldmath$\sigma$} 
 &=& 
 \sum_\alpha
 \varphi^{{\rm cl}\dag}_\alpha\sigma\varphi^{\rm cl}_\alpha ,
\end{eqnarray}
\end{subequations}
where $\rho$, $\tau$ and ${\bf J}$ are time-even while ${\bf j}$ and
{\boldmath$\sigma$} are time-odd.  Note that time-even and time-odd
terms appear in particular combinations, a feature which is crucial to
render the functional invariant under Galilean
transformations~\cite{engel}.

\subsection{Problems with $v/c$ expansions}
\label{sec:vcproblem}

Nonrelativistic limits are usually obtained by straightforward
expansion in orders $v/c$ or $p/m$, respectively, e.g. in the
Foldy-Wouthuysen transformation (\cite{Zuber}, for the nuclear case
see \cite{Thi86b,reinhard}). We briefly review the steps from
\cite{reinhard}.  One starts from the relativistic equations of
motion, for the present model derived from the Lagrangian
(\ref{eq:prmf_den})
\begin{subequations}
\begin{eqnarray}
  0
  &=&
  \left(
   {\rm i}\gamma^\mu\partial_\mu-m
   +S+\gamma^\mu V_\mu
  \right)\psi_\alpha
  ,
\\
  S
  &=&
  -c_s\varrho_s
  ,\quad
  V_\mu
  =
  c_v\varrho_\mu.
\end{eqnarray}
\end{subequations}
with the self-consistent scalar and vector potentials $S$ and $V_\mu$,
the latter decomposing as
\begin{equation}
 V_\mu
 =
 (V_0,-{\bf V})
 =
 (c_v\varrho_0,-c_v\mbox{\boldmath$\varrho$}).
\label{eq:Vpot}
\end{equation}
We insert the decomposition (\ref{eq:expa-wf}) and solve the lower-component equation for
$\varphi^{\rm(d)}$.  Keeping only terms up to order $p/m$
yields
\begin{subequations}
\label{eq:updw}
\begin{eqnarray}
  \varphi^{\rm(d)}
  &=&
  B_0\sigma\!\cdot\!(\hat{\bf p}-{\bf V})\varphi^{\rm(u)},
\label{eq:down}\\
  B_0
  &=&
  \frac{1}{2m+S-V_0} 
  \approx
  \frac{1}{2m} ,
\end{eqnarray}
\end{subequations}
where $\hat{\bf p}=-{\rm i}\nabla$. The approximation $B_0=1/m$ ignores
the density dependence in $B_0$. It suffices for the present studies. 
The form (\ref{eq:updw}) violates
the normalization (\ref{eq:normcond}) to second order in $p/m$. The
procedure of \cite{reinhard} restores (ortho-)normalization at
operator level by first imposing the relativistic
normalization (\ref{eq:normcond}) up to $(p/m)^2$
\begin{eqnarray}
  1
  &=&
  \int {\rm d}^3r\,
  \varphi^{{\rm(u)}\dag}
   \left[1+\hat{\cal T}\right]
  \varphi^{\rm(u)}
\nonumber\\
  \hat{\cal T}
  &=&
  B_0^2
  \left(\sigma\!\cdot\!({\bf p}-{\bf V})\right)^2
  \,.
\end{eqnarray} 
We introduce $\varphi^{\rm (cl)}$ so as to recover the
nonrelativistic normalization (\ref{eq:clnormcond}).
Thus we identify
\begin{eqnarray*}
  1
  &=&
  \int {\rm d}^3r\,
  \underbrace{\varphi^{{\rm(u)}\dag}
   \left[1+\hat{\cal T}\right]^{1/2}}_{\varphi^{\rm ((cl)\dag}}
  \underbrace{
   \left[1+\hat{\cal T}\right]^{1/2}
  \varphi^{\rm(u)}}_{\varphi^{\rm ((cl)}}
\end{eqnarray*} 
We expand in second order of $p/m$ and obtain
\begin{equation}
  \varphi^{\rm(u)}
  =
  \Big[1-\frac{1}{2}\hat{\cal T}\Big]\varphi^{\rm(cl)}
\label{eq:upcl}
\end{equation}
which, together with relation (\ref{eq:down}), provides a complete
description of $\psi$ in terms of $\varphi^{\rm(cl)}$ up to order
$p/m$. 
Using the relation
$
(\mbox{\boldmath$\hat\sigma$}\!\cdot\!{\bf A})
(\mbox{\boldmath$\hat\sigma$}\!\cdot\!{\bf B})
=
{\bf A}\!\cdot\!{\bf B}
-{\rm i}
{\bf A}\!\cdot\!(\mbox{\boldmath$\hat\sigma$}\!\times\!{\bf B})
$ yields
\begin{equation}
  \frac{\hat{\cal T}}{2B_0^2}
  =
  (\hat{\bf p}-{\bf V})^2
  +
  {\rm i}(\hat{\bf p}-{\bf V})
   \!\cdot\!\left(\mbox{\boldmath$\hat\sigma$}\!\times\!
  (\hat{\bf p}-{\bf V})\right).
\label{eq:explT}
\end{equation}
We insert eqs. (\ref{eq:Vpot}), (\ref{eq:upcl}) and (\ref{eq:explT})
into expressions (\ref{eq:relat-dens}) for the relativistic densities
and finally obtain up to order $(p/m)^2$
\begin{subequations}
\label{eq:vcdens}
\begin{eqnarray}
  \varrho_s
  &=&
  \rho
  -
  2B_0^2\Big[
   \tau\!-\!\nabla\!\cdot\!{\bf J}
    \!-\!
    c_v\mbox{\boldmath$\varrho$}\!\cdot\!(2{\bf j}
    \!+\!
    \nabla\!\times\!\mbox{\boldmath$\sigma$})
    \!+\!
    c_v^2\mbox{\boldmath$\varrho$}^2\rho
  \Big]\qquad
\\
  \varrho_0
  &=&
  \rho
  ,\quad
  \mbox{\boldmath$\varrho$}
  \;=\;
  2{{B}_0}
  \Big({\bf j}-c_v\mbox{\boldmath$\varrho$}\rho
  +\frac{1}{2}\nabla\!\times\!\mbox{\boldmath$\sigma$}\Big).
\end{eqnarray} 
\end{subequations}

The same relations are obtained when going through the
Foldy-Wouthuysen transformation up to second order. 
Note that the spatial part of the vector density
$\mbox{\boldmath$\varrho$}$ is at least of first order such that
the correction from $\hat{\cal T}$ would be of third order and
is discarded. 
The result is correct for stationary states where all time-odd
densities and fields vanish, i.e. where ${\bf j}=0$,
$\mbox{\boldmath$\sigma$}=0$, and ${\bf V}=0$.
The expressions (\ref{eq:vcdens}) when inserted into the
Lagrangian (\ref{eq:prmf_den}) produce a serious defect: the
emerging nonrelativistic Lagrangian is not Galilei invariant. The
Lorentz invariant scalar density $\rho_s$ does not translate to a
Galileian invariant expression and the same happens for the combination
$\varrho_0^2-\mbox{\boldmath$\varrho$}^2$.
Correct expressions should form the Galileian invariant
combinations $\rho\tau-{\bf j}^2$ and
$\rho{\nabla}\!\cdot\!{\bf J}
     +{\bf j}\!\cdot\!({\nabla}\!\times\!\mbox{\boldmath$\sigma$})$
as we will see later.  

In order to elucidate the problem, we consider the transformation
properties for the simple case of an explicit boost of the whole
system. Let us start with a well-checked situation,
a stationary state for which ${\bf j}=0$ and $\mbox{\boldmath$\sigma$}=0$. 
We boost the system by a constant velocity ${\bf u}$ (in units of $c$).
 All quantities in
the boosted system will be distinguished by a prime.  The
normalization (\ref{eq:normcond}) becomes in the boosted frame
\begin{equation}
  1
  =
  \int {\rm d}^3r'\,\varrho'_0
  =
  \int {\rm d}^3r\sqrt{1-{\bf u}^2}\,\frac{\varrho_0}{\sqrt{1-{\bf u}^2}}.
\label{eq:secondnorm}
\end{equation}
Note the transformation of the volume element exactly
countering that of the density $\varrho_0$. The volume
dilatation factor is negligible to order $u^1$ but contributes in
second order. The above nonrelativistic expansion to second
order had violating terms  at that  order. The
example shows that the mistake lies in neglecting a second order
correction of the volume element in the normalization condition.

\subsection{Map instead of expansion}

The previous discussion shows that a straightforward nonrelativistic
expansion with all kinetic contributions is consistent only up to
first order $p/m$ (or boost velocity $u$, respectively), while the
crucial relativistic corrections to a classical Schr\"odinger equation
emerge from second order terms, namely spin-orbit coupling and
effective-mass terms. These require a special handling of the
normalization condition like in the example of
Eq. (\ref{eq:secondnorm}). We thus leave the straightforward paths of
$p/m$ expansion to now aim at a generalized mapping of the
relativistic functional into a nonrelativistic one, trying to
incorporate all second-order effects.

The key point is accounting for the relativistic volume-element
compression as in Eq. (\ref{eq:secondnorm}). To deal with arbitrary
nonstationary situations we need to allow a boost velocity field. Thus
we modify the normalization condition (\ref{eq:normcond}) to display
explicitly the compression factor with respect to the local boost
velocity, which in turn is expressed in terms of the classical
densities and currents:
\begin{subequations}
\begin{eqnarray}
  1
  &=&
  \int {\rm d}^3r\sqrt{1-{\bf u}^2({\bf r})}\, \varrho_0
\\
  {\bf u}({\bf r})
  &=&\frac{\mbox{\boldmath$\varrho$}}{\varrho_0}\equiv
  \frac{2{B}_0}{\rho}
   \Big({\bf j}-{\bf V}\rho +\frac{1}{2}\nabla\!\times\!\mbox{\boldmath$\sigma$}\Big).
\label{eq:normvel}
\end{eqnarray}
The expansion (\ref{eq:upcl}) thus
is slightly modified to the map
\begin{eqnarray}
  \varphi^{\rm(u)}
  &=&
  \Big[1-\frac{1}{2}\hat{\cal T}\Big]\varphi^{\rm(cl)}
  (1+\frac{1}{4}{\bf u}^2)
\nonumber\\
  &\approx&
  \Big[1-\frac{1}{2}\hat{\cal T}+\frac{1}{4}{\bf u}^2\Big]
  \varphi^{\rm(cl)}.
\label{eq:upclvel}
\end{eqnarray}
\end{subequations}
Things now proceed as in section \ref{sec:vcproblem}, but
the term $\propto{\bf u}^2$ cancels the unwanted one in $\varrho_s$
and adds a desired one in $\varrho_0$. This now leads to the consistent result
\begin{subequations}
\label{eq:mapdens}
\begin{eqnarray}
  \varrho_s
  &=&
  \rho
  -
  \frac{2B_0^2}{\rho}
  \Big[
     \rho\tau-{\bf j}^2
    -\rho \nabla\!\cdot\!{\bf J}
    -{\bf j}\!\cdot\!(\nabla\!\times\!\mbox{\boldmath$\sigma$})
\nonumber\\
  &&\hspace*{4.4em}
    -\frac{1}{4}(\nabla\!\times\!\mbox{\boldmath$\sigma$})^2
  \Big]
\\
  \varrho_0
  &=&
  \rho
  +
  \frac{2B_0^2}{\rho}
  \Big({\bf j}-{\bf V}\rho+\frac{1}{2}\nabla\!\times\!\mbox{\boldmath$\sigma$}
       \Big)^2
  ,
\\
  \mbox{\boldmath$\varrho$}
  &=&
  2{{B}_0}
  \Big({\bf j}-{\bf V}\rho+\frac{1}{2}\nabla\!\times\!\mbox{\boldmath$\sigma$}
       \Big).
\end{eqnarray} 
\end{subequations}
The scalar density shows the wanted
Galilean-invariant combinations and the vector density reproduces the
correct invariance property,
$\varrho_\mu\varrho^\mu
  =
  \rho_0^2 -\mbox{\boldmath$\varrho$}^2
  \approx
  \rho^2$,
up to terms of second order, of course.
We thus insert the mapped scalar and vector densities
(\ref{eq:mapdens}) into the interaction Lagrangian density
(\ref{eq:prmf_den}), getting (up to second order)
\begin{eqnarray}
  {\cal L}_{\rm c}
  &=&
  \frac{c_s+c_v}{2}\rho^2
	-{2 c_s B_0^2}
  \Big[\rho\tau-{\bf j}^2
\nonumber\\
  &&-(\rho{\nabla}\!\cdot\!{\bf J}
     +{\bf j}\!\cdot\!({\nabla}\!\times\!\mbox{\boldmath$\sigma$}))
   -\frac{1}{4}(\nabla\!\times\!\mbox{\boldmath$\sigma$})^2
  \Big].
\label{eq:finfunc}
\end{eqnarray}
That result is manifestly Galilean invariant. 

It has the same form as the basic version of the Skyrme Hamiltonian
density \cite{Ben03aR} that is being employed for the description of
finite nuclei.  This comes as no surprise, since the covariant
Lagrangian that we started with, Eq. (\ref{eq:prmf_den}), consists of
the basic terms of the Lagrangian of the point-coupling variant of the
RMF model for nuclear structure, RMF-PC \cite{buervenich}.  Both SHF
and RMF-PC are formulated in terms of point couplings of spinors and
thus display this close relationship. This relation between covariant
models and their Galilean-invariant counterparts is of importance when
one analyzes, for example, spin excitation mechanisms in
nonrelativistic time-dependent Hartree Fock employing the Skyrme
functional \cite{spin}.

Compared with the Skyrme functional, however, there is one
additional term $\propto(\nabla\!\times\!\mbox{\boldmath$\sigma$})^2$ 
which is an allowed term in the  Skyrme functional, but usually
neglected.
There is, however, no gradient term $\propto\rho\Delta\rho$ which is mandatory for
the description of finite-size systems.  That is no surprise because we had
started from a simplified Lagrangian without gradient terms. The more complete model would also include terms
such as $\rho_s\Delta\rho_s$ and $\rho_\mu\Delta\rho^\mu$, whose expansion proceeds quite similarly
and in the nonrelativistic limit yields gradient terms.
There is a subtle
difference, though: the nonrelativistic mapping would also produce gradient kinetic terms like
$\rho\Delta\tau$, which are neglected assuming that the gradient correction as such is small and
second order relativistic corrections to it are negligible.  The counter argument is that there are
two quite different notions of smallness involved here that may not be combined.

\section{Conclusion}
We have studied the nonrelativistic limit of a self-consistent relativistic theory
with the aim of recovering a minimum of relativistic effects, the
spin-orbit force, together with Galilean invariance in the resulting
nonrelativistic theory.
Note that Galilean invariance requires keeping all time-odd terms, which play a crucial role in the
formulation of dynamics. This applies to the spatial components of the
relativistic vector density as well as to the current and spin densities in the
nonrelativistic domain. Previous derivations rarely studied the full
dynamical case. 

The procedure started out with a straightforward $v/c$ expansion, which encountered 
inconsistencies, because the spin-orbit term appears only
in second order of $v/c$ while Galilean invariance comes
out correctly only to first order. The key finding is that a strictly
nonrelativistic
theory is not easily
compatible with the appearance of a spin-orbit term. To be more
precise: {\em for relativistic effective field theories, based on density functionals, 
it is not possible to derive a sufficiently complete
nonrelativistic theory by mere expansion and order counting}. Instead
the more general concept of a nonrelativistic mapping is needed,
namely to derive an effective nonrelativistic theory which includes
as many features of the given relativistic theory as desired.
Starting from the simple consideration of Lorentz contraction
of the spatial volume element, we have derived such a mapping for a
covariant self-consistent model. This mapping manages to provide a manifestly Galilean invariant
theory which correctly incorporates the spin-orbit and effective-mass
terms and, if one starts with the RMF-PC model in nuclear physics,  
merges into the widely used Skyrme-Hartree-Fock approach when
neglecting the involved density dependences of the spin-orbit and
effective-mass term. Extensions of the scheme developed here are
in progress.

\section*{Acknowledgement}
The work was supported by the BMBF (06 ER 124), DFG, CONACyT and
DGAPA. Helpful discussions with M. Thies (Erlangen) are
gratefully acknowledged.
P.O.H. thanks also the kind hospitality at
the {\it Frankfurt Institute for Advanced Studies} during which these
results were obtained, A. S. is grateful for the kind hospitality
during his stay at the {\it Institute for Theoretical Physics} at the University of Frankfurt,
where these results were obtained.

\end{document}